\documentclass[prd,preprintnumbers,amsmath,amssymb,a4paper,nofootinbib,showpacs]{revtex4}

\usepackage{graphicx}
\usepackage[utf8]{inputenc}

\begin{document}

\markboth{}
{\textit{D.A. Fagundes, M.J. Menon, P.V.R.G. Silva}}

\title{Total Hadronic Cross Section Data and the Froissart-Martin Bound}

\author{D.A. Fagundes, M.J. Menon, P.V.R.G. Silva}

\affiliation{Universidade Estadual de Campinas - UNICAMP\\
Instituto de F\'{\i}sica Gleb Wataghin \\
13083-859 Campinas, SP, Brazil \\
fagundes@ifi.unicamp.br, menon@ifi.unicamp.br, precchia@ifi.unicamp.br}


\begin{abstract}
The energy dependence of the total hadronic cross section at high energies is investigated
with focus on the recent experimental result by the TOTEM Collaboration at 7 TeV
and the Froissart-Martin bound.
On the basis of a class of analytical parametrization
with the exponent $\gamma$ in the leading logarithm 
contribution as a free parameter, different variants of fits to 
$pp$ and $\bar{p}p$ total cross section data above 5 GeV are developed.
Two ensembles are considered, the first comprising data up to 1.8 TeV,
the second also including the data collected at 7 TeV. We shown that in all fit variants
applied to the first ensemble
the exponent is statistically consistent with $\gamma$ = 2. Applied to the second ensemble, however, 
the same variants yield $\gamma$'s above 2, a result
already obtained in two other analysis, by U. Amaldi \textit{et al}. and by the UA4/2 Collaboration.
As recently discussed by Ya. I. Azimov,
this faster-than-squared-logarithm rise does not necessarily violate unitarity.
Our results suggest that the energy dependence of the hadronic total cross section at high energies 
still constitute an open problem.
\end{abstract}

\maketitle

\vspace{0.7cm}

\begin{center}
\textit{To appear in the Brazilian Journal of Physics, 2012}
\end{center}

\vspace{0.7cm}

\textbf{Table of Contents}

\vspace{0.2cm}

I. Introduction

\ \ \ \ \ I.A General Aspects

\ \ \ \ \ I.B Purposes and Goals of the Paper

II. Analytical Parametrization and Data Ensemble

\ \ \ \ \ II.A Analytical Parametrization and Previous Results

\ \ \ \ \ II.B Data Ensemble and Critical Comments

III. Fitting Procedure, Variants and Results

\ \ \ \ \ III.A Method 1

\ \ \ \ \ \ \ \ III.A.1 Results for the $\sqrt{s}_{max}$ = 1.8 TeV Ensemble

\ \ \ \ \ \ \ \ III.A.2 Results for the $\sqrt{s}_{max}$ = 7 TeV Ensemble

\ \ \ \ \ III.B Method 2

\ \ \ \ \ \ \ \ III.B.1 Results for the $\sqrt{s}_{max}$ = 1.8 TeV Ensemble

\ \ \ \ \ \ \ \ III.B.2 Results for the $\sqrt{s}_{max}$ = 7 TeV Ensemble

IV. Conclusions and Final Remarks


Appendix A: Predictions for the Ratio Between the Real and Imaginary Parts of the 
Forward Amplitude

\vspace{0.5cm}


\newpage

\section{Introduction}
\label{sec:1_intro}

\subsection{General Aspects}
\label{subsec:1.1}

High-energy particle collisions constitute the main experimental tool in the investigation of the inner structure of matter. In Particle Physics, high-energy usually means center-of-mass energies above 10 $m_p$ $\sim$ 10 GeV, where $m_p$ is the proton mass\footnote{As usual in high-energy physics we adopt the system of units
in which $c$ = $\hbar$ = 1. Typical units of energy are 1 GeV = $10^{9}$ eV and 1 TeV = $10^{12}$ eV.} \cite{perkins}.
Presently, for particle-particle and antiparticle-particle collisions, the highest energies reached in accelerators concern proton-proton ($pp$) and antiproton-proton ($\bar{p}p$) interactions, corresponding to 7 TeV and $\sim$ 2 TeV, respectively. These hadronic processes are expected to be described by the Quantum Chromodynamics (QCD), the
non-Abelian gauge field theory of the strong interactions \cite{qcd}.

As a non-Abelian theory, the gluons (the field quanta) themselves carry a color charge and can
therefore interact with other gluons\footnote{This property contrasts with QED since photons do not have electric charge.}.
The dynamical consequence is a running coupling constant $\alpha_s$: the color charge is small at short distances
and large at large distances, leading to two different regimes named asymptotic freedom and confinement,
respectively \cite{perkins,qcd,halzen}. In hadron-hadron collisions these regimes correspond
to two sectors that have been known as hard scattering (small distances and large values of the momentum transfer) 
and soft scattering (large distances and small values of the momentum transfer). The confinement
barrier, where $\alpha_s \sim$ 1, is typical of distances of the order of
1 fm and therefore ``peripheral" hadronic collisions correspond to the soft sector.

The great triumph of QCD concerns the perturbative techniques, successfully applied in the hard sector \cite{pqcd}. By contrast, soft interactions, characterized by small values of the momentum transfer, can not be treated through 
these techniques due to the rise of the  coupling constant as the momentum transfer decreases. As a consequence, soft physics demands first principles and nonperturbative approaches, which means the non-trivial investigation of the vacuum structure, intricate Monte Carlo simulations and complex analytic formalisms \cite{npqcd}.
However, despite the  success of nonperturbative QCD  in the investigation
of the static hadronic properties (bound states), a formal approach to
\textit{soft scattering states}, based on first principles and without model assumptions, is still missing \cite{pred,land}
and that implies in some fundamental problems.

Soft scattering embodies elastic collisions and diffraction dissociation (single and double) \cite{pred} and here comes one of the striking features of QCD:
elastic scattering, the simplest kinematic collision process, just constitute one 
of the greatest dynamic problems for the theory of the strong interactions.
In this respect, the Optical Theorem \cite{merz} plays a crucial role since it connects the forward \textit{elastic} scattering
amplitude with the most important physical
quantity characterizing a collision process, namely the total cross section \cite{pred,land}.
Therefore, the lack of a pure QCD result for the elastic amplitude puts serious limits in the 
theoretical investigation
of the total hadronic cross section. On the other hand, and more important for our purposes, experimental information on the behavior of the total cross section may, in principle, be used as input providing new insights in the development
of the theory in the soft sector (the inverse problem), at least in what concerns the forward elastic
scattering amplitude.

 Operationally the total cross section
is defined by \cite{pred,matthiae}
\begin{equation}
\sigma_{tot} = \frac{N_{el} + N_{inel}}{\mathcal{L}},
\nonumber
\end{equation}
where $\mathcal{L}$ is the luminosity (flux per unit area) and 
$N_{el}$,  $N_{inel}$ are the rate of elastic and inelastic interactions,
respectively
(scattered fluxes). From the definition, two interpretations
emerges for $\sigma_{tot}$, one statistical, as a probability of interaction (ratio between incident and scattered particles) and another
geometrical, as an effective area of interaction, usually measured in $mb$
for hadronic scattering.

Experiments indicate that
at lower energies, below $\sim$ 2 GeV, $\sigma_{tot}$ is characterized by narrow peaks, caused by the formation of resonances (bound states). As the energy increases,
reaching the  scattering region, $\sigma_{tot}$ decreases very slowly up to $\sim$ 20 GeV and then starts to grow smoothly and monotonically (without bumps or dips), as illustrated by the experimental points in
Figure \ref{fig:1}, in the case of $pp$ and $\bar{p}p$ scattering (see also Figure 41.10 in \cite{pdg2010}).
The energy dependence of  $\sigma_{tot}$  is therefore a crucial point
since, above the resonance region, it is directly related to the soft sector (optical theorem)
giving information on the dynamics of the elastic interaction \cite{pred,land}.

\subsection{Purposes and Goals of the Paper}
\label{subsec:1.2}

Although the rise of the total hadron-hadron cross sections
at high energies is an experimental fact, the exact energy dependence
involved has been a long-standing problem.
Several phenomenological models, with distinct physical pictures and good descriptions
of the available data,
make different asymptotic predictions \cite{matthiae,kaspar,fiore}
and the only general, formal, widely
accepted result is the well-known Froissart-Martin (FM) bound 
\cite{froissart,martin}
\begin{equation}
\sigma_{tot}(s) \leq c\, \ln^2 \frac{s}{s_0},
\label{eq:1}
\end{equation}
where $s$ is the squared center-of-mass energy, 
$s_0$ a constant,
and a bound on the first term on the right-hand side,
\begin{equation}
c \leq \pi/m_{\pi}^2 \approx  60\ \mathrm{ mb},
\label{eq:2}
\end{equation}
has been obtained by Lukaszuk and Martin \cite{lukmar}. Even recently, the foundations of this 
key result in soft hadronic physics have been discussed in
the literature \cite{martin09,wmrs,azimov}. In particular, Azimov has presented
a short critical review on the
assumptions supporting the derivation of the bound \cite{azimov}.

Even depending on the unknown squared mass scale $s_0$, the numerical values associated with
the FM bound lie far above the existing data for the total cross section.
For example, for $s_0$ in the interval
1 - 50 GeV$^2$, the bound in (\ref{eq:1}-\ref{eq:2}) at 1 TeV is of order of 10 b, much larger than
the  100 mb cross sections typically found in experiments at the highest energies. 
However, the bound also comprises a maximum rate of rise for the cross section
with the energy, namely the squared logarithm  behavior at the asymptotic energy region
and that is the point we are interested in here.
Different dependencies have been extensively tested and discussed in the literature, 
specially in the context of amplitude analysis, which are characterized by analytical parametrization for the
total cross section and fits to forward data. The most common functional forms 
employed in the highest-energy region include different combinations of constant, linear/quadratic logarithm
dependencies and $s$ power laws, as discussed, for example, in 
\cite{ckk,lm,igi1,igi2,igi3,compete1,compete2,bh1,bh2,bh3} and references therein.
Nonetheless, like the case of specific phenomenological models \cite{kaspar,fiore,matthiae}, these analyses present good descriptions of the available data with different functional forms 
and hence offer different physical pictures.

New results from the CERN Large Hadron Collider (LHC)
and new estimates from the Pierre Auger Observatory for the
proton-proton total cross sections
are expected to shed light
on the subject, not only by selecting phenomenological models/pictures, but also
by providing information on the
degree of possible saturation of the bound in terms of the energy dependence of
the cross section.
In fact, at 7 TeV the first result for the total
cross section by the TOTEM\footnote{TOTal Elastic and diffractive cross section Measurement} Collaboration, a luminosity-dependent measurement \cite{totemsigma},
indicates consistency with a $\ln^2 s$ dependence, as
predicted ten years ago by the parametrization that was ranked highest by the COMPETE\footnote{Computerized Models, Parameter Evaluation for Theory and Experiment} Collaboration
\cite{compete1,compete2}, also quoted in the Review of Particle Physics
by the Particle Data Group \cite{pdg2010}. Therefore, these results favor 
the  maximum increase rate allowed by the bound.

By contrast, at least two previous almost-model-independent analyses, based on data ensembles at
lower energies, have indicated that the exponent in the logarithm
may be somewhat larger than 2 \cite{amaldi,ua42}. 
Moreover, 
on the theoretical side,
as recently discussed by Azimov \cite{azimov}, it is not obvious if
Martin's derivation, in the context of axiomatic local quantum field 
theory, can be directly applied to hadronic processes. Formal arguments suggest
that the total cross section could grow faster than $\ln^2{s}$. 
A recent result from lattice QCD \cite{meg}
indicates an universal asymptotic squared logarithm  dependence for the hadronic total cross 
section. That conclusion nonetheless rests on
specific assumptions and is hence neither unique or exclusive.

In view of these facts the following question arises.
Taking into account the TOTEM result and considering the exponent in the logarithm
as a free fit parameter, the $\ln^2{s}$ dependence is in fact a unique solution
describing the asymptotic rise of the total cross section, or the data can be
statistically described by another solution, rising faster (or slower) than 
$\ln^2{s}$? To answer this question is the goal of this work.

For this purpose, we shall revisit the analytical pa\-ram\-e\-tri\-za\-tion introduced by
Amaldi \textit{et al}. in the seventies \cite{amaldi}, also employed by the UA4/2 Collaboration,
in the nineties \cite{ua42}, characterized by Reggeon contributions at low
energies and a leading contribution at high energies parametrized by 
a power law in $\ln{s}$, the real exponent $\gamma$ as a free parameter.
As will be recalled along the text, the two analyses cover different energy intervals
and led to an exponent exceeding 2. For future reference, we note that
the above-mentioned highest-ranking parametrization, obtained by the COMPETE Collaboration \cite{compete1,compete2},
can be regarded as a particular instance of this parametrization in which
the exponent was fixed at $\gamma$ = 2.

All these works developed simultaneous fits to $\sigma_{tot}$ and the $\rho$ parameter
(ratio between the real and imaginary parts of the forward amplitude).
Here, for reasons to be discussed in detail, we shall analize only
the $\sigma_{tot}$ data, from $pp$ and $\bar{p}p$ scattering above
$\sqrt{s}$ = 5 GeV. In order to investigate the effect of the recent experimental result
for the $pp$ total cross sections by the TOTEM Collaboration, we first consider
all the previous existing data, covering the region up to
1.8 TeV, and then add to the ensemble the data at 7 TeV. We choose two different initial values for the
free fit parameters and six fitting variants. 
We show  that, with data up to 1.8 TeV, the real exponent
in the logarithm term is statistically consistent with $\gamma$ = 2,  
as predicted by the COMPETE Collaboration. However, with the addition of the data
at 7 TeV, the fits indicate exponents larger than 2 in all cases investigated.

The manuscript is organized as follows. In Sect. \ref{sec:2} we introduce the analytical parametrization,
followed by some comments on its applicability
and a critical discussion on the selected data ensemble. In Sect. \ref{sec:3} we present the fit procedures,
variants and results.
The conclusions and some final remarks 
are the contents of Sect. \ref{sec:4}.

\section{Analytical Parametrization and Data Ensemble}
\label{sec:2}
\subsection{Analytical Parametrization and Previous Results}
\label{subsec:2.1}

The class of analytical parametrization \cite{amaldi,ua42} consists of
two components for the total hadronic cross section
associated with low- ($L$) and high-energy ($H$) contributions:
\begin{equation}
\sigma_{tot}(s) = \sigma_{LE}(s) +  \sigma_{HE}(s).
\label{eq:3}
\end{equation}
The first term accounts for the decreasing 
of the total cross section and
the differences
between particle-particle and particle-antiparticle scattering at low energies and is expressed by
\begin{equation}
\sigma_{LE}(s) = a_1\, \left[\frac{s}{s_l}\right]^{-b_1} + 
\tau\, a_2\, \left[\frac{s}{s_l}\right]^{-b_2},
\label{eq:4}
\end{equation}
where $s_l$ = 1 GeV$^2$ (fixed) while $a_1$, $b_1$, $a_2$, $b_2$ are free fit parameters, and 
\begin{equation}\nonumber
 \tau =\left\{
\begin{array}{rl}
 -1 & \text{for } \text{particle-particle} \\
 +1 & \text{for } \text{antiparticle-particle.}
\end{array}\right.
\end{equation}
The  second term accounts for the rising of the cross section at higher energies and is given by
\begin{equation}
\sigma_{HE}(s) =
\alpha\, + \beta\, \ln^\gamma \frac{s}{s_h},
\label{eq:5}
\end{equation}
where $\alpha$, $\beta$, $\gamma$, $s_h$ are real free parameters.
For further discussion, let us briefly recall some formal aspects and previous results
associated with this class of parametrization. 

In the context of the Regge-Gribov theory all terms in Eqs. (\ref{eq:4}) and (\ref{eq:5})
have specific physical interpretations, namely Reggeon and Pomeron exchanges at
low and high energy regions, respectively \cite{compete2}. The Reggeons correspond to mesons resonances
families with the adequate quantum numbers in the $t$-channel process
and represented by trajectories interpolating the data on plots
of spin $J$ versus the square of their masses (Chew-Frautschi plot). In this case,
$b_1$ and $b_2$ correspond to the intercept of the trajectories and
$a_1$, $a_2$ the Reggeon strengths (residues). For $pp$ and $\bar{p}p$ scattering,
$\sigma_{LE}$ is associated with two Reggeons, the first one with $C$ = + 1
($a_ 2$ and $f_2$ mesons trajectories) and the second with 
$C$ = - 1 ($\rho$ and $\omega$ mesons trajectories), corresponding to $\tau$ = + 1
and $\tau$ = -1, respectively. The type of Pomeron contribution depends on the $\gamma$ value.
For $\gamma$ = 1, the constant plus $\ln s$ terms correspond to a double pole at $J$ = 1
and for $\gamma$ = 2 a triple pole (expressed by $\ln^2 s$, $\ln s$ and the constant terms).
Up to our knowledge, the case of real exponent and 0 $< \gamma <$ 2 corresponds
to a strong-coupling scenario (critical Pomeron) \cite{gm,rmk}
and a fractional power, $\gamma = 3/2$ (in general $1<\gamma<2$), is indicated by the mini-jet QCD model with infrared gluon resummation \cite{giu1,giu2}.

As commented before, the above parametrization has been introduced 
by Amaldi \textit{et al}. \cite{amaldi}, with $\sigma_{LE}$  expressed
as function of the lab energy, $\sigma_{HE}$ as function of $s$ and
with a fixed scale constant $s_{h}$ = 1 GeV$^2$. Simultaneous fits to $\sigma_{tot}$ and $\rho$ data
(via dispersion relations) from $pp$ and $\bar{p}p$ scattering in the
interval 5 $< \sqrt{s} \leq $ 62 GeV has lead to the result
\vspace{0.2cm}

\centerline{$\gamma$ = 2.10 $\pm$ 0.10.}

\vspace{0.2cm}

The same functional form was subsequently used by the UA4/2 Collaboration \cite{ua42}.
Simultaneous fits to $\sigma_{tot}$ and $\rho$ data
from $pp$ and $\bar{p}p$ scattering in a larger interval,
5 $< \sqrt{s} \leq $  546 GeV, have yielded the result 

\vspace{0.2cm}

\centerline{$\gamma$ = 2.25$^{+0.35}_{-0.31}$.}

\vspace{0.2cm}

More recently the COMPETE Collaboration has carried out a detailed and extensive study on 
possible analytic parametrizations
including all measured $\sigma_{tot}$ and $\rho$ data from $pp$, $\bar{p}p$, 
meson-$p$ scattering, among other
processes, at $\sqrt{s} \geq$ 4 GeV and $\bar{p}p$ data up to
1.8 TeV \cite{compete1,compete2}.
Different aspects of fit qualities have been considered in a ranking procedure with the same
$\sigma_{LE}$ structure and $\sigma_{HE}$ parametrized either with $\gamma$ = 1 or $\gamma$ = 2
and the constant term. The parametrization with $\gamma$ = 2 was ranked highest.
As commented before, the predictions from this analysis, carried out 10 years ago, 
indicate consistency with the TOTEM result at 7 TeV and therefore agrees with
the saturation of the squared logarithm dependence in the FM bound.

\subsection{Data Ensemble and Critical Comments}
\label{subsec:2.2}

Two aspects of our choice of data ensemble deserve special atention. One aspect
concerns the reactions and the other the physical quantities to be investigated,
as discussed in what follows.

First, the squared logarithm dependence in the FM bound is an asymptotic result, and
the parametrization (\ref{eq:3}-\ref{eq:5}) covers particle-particle and particle-antiparticle interactions.
For those reasons we shall consider
only the cases with the \textit{highest energy interval in terms of available data}, namely $pp$ and $\bar{p}p$ scattering.
With this restrictive choice we do not take account of any constraint dictated by a
supposed universal behavior, or data from other reactions in the region
of intermediated and low energies, as for example the meson-proton cases.
Following Amaldi \textit{et al}. and the UA4/2 Collaboration,
we focus our analysis on data at $\sqrt{s} \geq$ 5 GeV$^2$.

Second, as we have commented before, amplitude analyses of the growth of the total cross section include
information on the $\rho$ parameter, through dispersion relations (integral and/or
derivatives forms), or the asymptotic prescriptions for crossing even and odd
amplitudes (Phragm\'en-Lindel\"off theorem) \cite{eden,block1,block2}. 
With the exponent $\gamma$ as a real free fitting parameter, the integral dispersion relations 
demand numerical
methods and therefore require an specific approach for error propagation from the uncertainty
in the $\gamma$ parameter. 
The use of prescriptions seem to us unjustified in the region of
intermediate and low energy data, since they are asymptotic results \cite{eden}.
By constrast, derivative dispersion relations allow an analytical approach 
and can be extended down to 4 - 5 GeV \cite{cudell1,cudell2} or even below (above the
physical threshold) 
in the form of a double infinite series \cite{am1,am2} or a single series \cite{fs}.
However, and that is a crucial point in this work, we shall not consider simultaneus fits
to total cross section data and  $\rho$ information 
for the six reasons that follows.

\begin{enumerate}

\item 
Strictly speaking the $\rho$ parameter is not a quantity with the same 
physical status as the total cross section since, in practice, it is evaluated as a free fit parameter in the Coulomb-nuclear interference region or inferred from analytical parametrization \cite{block1,block2}.

\item 
As the energy grows, it becomes progressively more difficult to determine 
$\rho$ \cite{block1,block2} and therefore the associated uncertainty, as can be easily
seen in plots of this parameter in terms of the energy. Even the COMPETE Collaboration refers
to the difficulty to adequately fit the $\rho$ data from $pp$ scattering \cite{compete2}.

\item 
Simultaneous fits to total cross section data and $\rho$ parameter
demand the use of dispersion relations with one subtraction \cite{disprel1,disprel2}
and therefore one more parameter, the subtraction constant. However
this constant does not have a physical interpretation, constrasting therefore 
with all the parameters in the $\sigma_{tot}$ parametrization (shortly discussed 
in Section \ref{subsec:2.1}).

\item 
In data reductions, the correlation of the subtraction constant
with all the other (physical) parameters affects the
fit results at both low- and high-energy regions, as discussed
in detail in \cite{alm}. As a consequence, the presence or not
of the subtraction constant may lead to different results.
In this respect we note that, although referred to in \cite{amaldi} the
value of the subtraction constant is not given (null or neglected?), in \cite{ua42} 
its fit value is $-$ 57 $\pm$ 4 and it is neglected in the prescriptions
or derivative dispersion relations used by the COMPETE Collaboration \cite{compete1,compete2}.
It should also be noted that even the prediction of the $\rho$
parameter from fits to total cross section data is affected by the
presence or not of the subtraction constant \cite{alm,alm01}.

\item 
As recently demonstrated by Ferreira \cite{erasmo1} and collaborators \cite{erasmo2,erasmo3}
(see also \cite{bsw}), the different slopes
associated with the real and imaginary parts of the hadronic amplitude in
the Coulomb-nuclear interference region, affect the extracted value of the $\rho$
parameter. This effect, however, has never been considered in the
experimental procedures used in the $\rho$ determination. The limited validity
of the relative phase between Coulomb and nuclear amplitudes, used in the
experimental procedures to determine $\rho$,
has also been discussed by Kundr\'at, Lokaj\'{\i}\v{c}ek and Vrko\v{c} \cite{klv}.

\item 
As a consequence of the above mentioned effects and the increasing uncertainty
in the $\rho$ determination with the energy, any possible deviation from an analytical 
parametrization for the total cross section, dictated by the experimental data at the 
highest energies, may be hidden or lost.
In other words, the inclusion of the $\rho$ information may erroneously anchor the rise
of the total cross section at the highest energies.

\end{enumerate}

These six critical points suggest that, at the high-energy region, fits restricted to
the total cross section data may avoid the bias introduced by both the $\rho$
parameter and the subtraction constant embodied in the dispersion relations. 
Therefore, although not being a usual procedure in amplitude analysis,
we understand that to explore the possibility of
focusing our fits only on the total cross section data constitute, at least, a
valid strategy.
For that reason  our data sample comprises only the $pp$ and $\bar{p}p$ total cross
section data above 5 GeV \cite{pdg2010}, including in special, the recent TOTEM result
at 7 TeV \cite{totemsigma}.

Two points must be stressed in the analysis that follows: [i] we are concerned only
with the rate of increase of the total cross section, not with the numerical value of the
FM bound; 
[ii] our data ensemble is only a subset of the data employed in the global analysis by the 
COMPETE Collaboration.

\section{Fittting Procedure, Variants and Results}
\label{sec:3}

By means of a luminosity-dependent method the TOTEM Collaboration has recently obtained
for the $pp$ total cross section at 7 TeV the value 
$\sigma_{tot}^{pp}$ = 98.3 $\pm$ 0.2$^{\mathrm{stat}}$ $\pm$ 2.8$^{\mathrm{syst}}$ mb
\cite{totemsigma}.
Our point here is to investigate the effect of this new data
in the previous analytical fits covering the energy interval
5 GeV $\leq \sqrt{s} \leq$ 1.8 TeV, as was the case with the
COMPETE Collaboration.  Based on the results by Amaldi \textit{et al}. and by the UA4/2
Collaboration, our strategy amounts to using the parametrization (\ref{eq:3}-\ref{eq:5}) to
investigate possible deviations from $\gamma$ = 2. 

However, several aspects related to fitting procedure and the alternatives allowed
by physical considerations or assumptions
deserve investigation. The three main points concern: (1) the nonlinearity of the fit demands
a methodology for the choice of the initial (feedback) values of the free parameters \cite{bev}; 
(2) as commented in Sect. \ref{subsec:2.1},
in the theoretical context the intercepts $b_1$ and $b_2$ are expected to be consistent
with spectroscopic data (Chew-Frautschi plot); (3) fixing $b_1$ and $b_2$ affects the fit results in both
low- and high-energy regions, due to the correlation among the free parameters.
In order to address these points the following methodology and fit variants have been considered.

In all cases under study we first consider the ensemble in the interval 5 GeV $\leq \sqrt{s} \leq $  1.8 TeV
and after that we include the data at 7 TeV from the TOTEM Collaboration. For
easier reference we will denote these
two ensembles by $\sqrt{s}_{max}$ = 1.8 TeV and $\sqrt{s}_{max}$ = 7 TeV, respectively.
In order to investigate the effects of different choices for the initial values
of the free parameters, we have followed two alternative procedures, named Method 1 and 
Method 2:

\vspace{0.2cm}

\noindent\emph{Method 1.}

Since the highest rank result by the COMPETE Collaboration
predicts a $\ln^2{s}$ dependence and is consistent with the TOTEM result at
7 TeV, we consider as a kind of orthodox choice 
to initialize our parametric set with
the central values they have
obtained in the simultaneous fit to $\sigma_{tot}$ and $\rho$ data, which includes $pp$, $\bar{p}p$
in the interval 5 GeV $\leq \sqrt{s} \leq $  1.8 TeV as a subset.
The numerical values \cite{pdg2010} are displayed in the second column of Table \ref{tab:1}.

\vspace{0.2cm}

\noindent\emph{Method 2.} 

Alternatively we have chosen $b_1$ = $b_2$ = 0.5
(average values for Reggeon intercepts), $\gamma$ = 2,
$a_1$ = $a_2$ = $\alpha$ = 1 mb, $s_h$ = 1 GeV$^2$ and
$\beta$ = 50 mb (simulating a saturation of the Lukaszuk-Martin bound \cite{lm}).
The numerical values are displayed in the first column of Table \ref{tab:3}.


The data reductions were carried out with the objects of the class TMinuit of 
the ROOT Framework \cite{root}. A Confidence Level (CL) of $\approx$ 68 $\%$ (one standard deviation),
was adopted in all fits, so that
the projection of the $\chi^{2}$ distribution in $(N+1)$-dimensional space ($N =$ number of free fit parameters) 
has the probability of 68 $\%$ \cite{bev}.

With both methods and the two ensembles, different variants were also considered. 
Sections \ref{subsec:3.1} and \ref{subsec:3.2} describe the variants and present the results obtained
with Methods 1 and 2, respectively, the discussion of the results being deferred to Sect. \ref{sec:4}.

\subsection{Method 1}
\label{subsec:3.1}

\subsubsection{Results for the $\sqrt{s}_{max}$ = 1.8 TeV Ensemble}
\label{subsubsec:3.1.1}

We consider the cases and notation that follows. The numerical results and statistical
information are displayed in Table \ref{tab:1}.

\begin{description}

\item[\bf{Direct Fit}:]

Initialized with the COMPETE parameters and $\gamma$ = 2 (fixed). The first run of the MINUIT Code 
yields the $\chi^2$ for that ensemble (second column in Table \ref{tab:1}) and the final run gives the 
direct fit result for that  ensemble (third column in Table \ref{tab:1}).

\item[\bf{Variant 1 (V1)}:]

Initialized with the Direct Fit parameters but now with the
exponent $\gamma$ as a free parameter (fourth column in Table \ref{tab:1}).

\item[\bf{Variant 2 (V2)}:]

Also initialized with the Direct Fit results. In this case, we consider $\gamma$ = 2 (fixed)
and $b_1$ and $b_2$ also fixed to the intercepts extracted from the spectroscopic data
on the $a_ 2$/$f_2$ and $\rho$/$\omega$ mesons trajectories, obtained by Luna, Menon and Montanha,
namely $b_1$ = 0.452 and $b_2$ = 0.558 \cite{lmm} (fifth column in Table \ref{tab:1}).

\item[\bf{Variant 3 (V3)}:]

Also initialized with the Direct Fit results. In this case we consider $\gamma$ free
and $b_1$ and $b_2$ fixed as in Variant 2 (sixth column in Table \ref{tab:1}).

\end{description}

In terms of the initial values and data reductions the following diagram summarizes the fitting
procedures:

\begin{equation}\nonumber
\text{COMPETE\ results} \rightarrow \text{Direct Fit} \rightarrow
\left\{\begin{array}{rl}
\text{Variant\ 1} \\
\text{Variant\ 2} \\
\text{Variant\ 3}
\end{array}\right.
\end{equation}

\subsubsection{Results for the $\sqrt{s}_{max}$ = 7 TeV Ensemble}
\label{subsubsec:3.1.2}

Here we consider as initial values each one of the corresponding results obtained
with the $\sqrt{s}_{max}$ = 1.8 TeV ensemble, namely those listed in Table \ref{tab:1}. 
The numerical results and statistical
information are displayed in Table \ref{tab:2}.

\begin{table}[ht]
\centering
 \caption{Results of Method 1 for the $\sqrt{s}_{max}$ = 1.8 TeV ensemble. Fit result from the 
COMPETE parameters as initial
values, Direct Fit and Variants 1, 2 and 3 (initial values from the Direct Fit result), together with the statistical
information: degrees of freedom (DOF) and reduced $\chi^2$. The parameters $a_1$, $a_2$, $\alpha$ and $\beta$
are in mb, $s_h$ is in GeV$^{2}$ and $b_1$, $b_2$ and $\gamma$ are dimensionless. }
 \label{tab:1}
 \begin{tabular}{cccccc}
\hline\noalign{\smallskip}
& Initial Values & Direct Fit & V1 & V2 & V3  \\
 & (COMPETE)      &            &    &    &    \\
\noalign{\smallskip}\hline\noalign{\smallskip}
$a_1$ & 42.53$\pm$0.23 & 54.6$\pm$4.0 & 54.6$\pm$1.7 & 55.4$\pm$3.1 & 57.0$\pm$2.5  \\
$b_1$ & 0.458$\pm$0.017 & 0.491$\pm$0.067 & 0.4907$\pm$0.0095 & 0.452 (fixed) & 0.452 (fixed)  \\
$a_2$ & 33.34$\pm$0.033 & 33.1$\pm$2.3 & 33.1$\pm$1.7 & 35.78$\pm$0.36 & 35.78$\pm$0.39 \\
$b_2$ & 0.545$\pm$0.007 & 0.540$\pm$0.016 & 0.540$\pm$0.012 & 0.558 (fixed) & 0.558 (fixed) \\
$\alpha$ & 35.45$\pm$0.48 & 34.2$\pm$2.5 & 34.19$\pm$0.24 & 32.14$\pm$0.99 & 31.38$\pm$0.98  \\
$\beta$ & 0.308$\pm$0.010 & 0.264$\pm$0.029 & 0.263$\pm$0.018 & 0.245$\pm$0.018 & 0.180$\pm$0.062  \\
$\gamma$ & 2  & 2 (fixed) & 2.001$\pm$0.026 & 2 (fixed) & 2.083$\pm$0.095  \\
$s_h$ & 28.9$\pm$5.4 & 12$\pm$12 & 12.2$\pm$1.5 & 5.7$\pm$3.0 & 2.8$\pm$2.6\\ \hline
DOF & 156 & 156 & 155 & 158 & 157 \\
$\chi^2/$DOF & 1.02 & 0.931 & 0.937 & 0.929 & 0.934  \\
\hline\noalign{\smallskip}
\end{tabular}
\end{table}

\begin{table}[ht]
 \centering
 \caption{Results of Method 1 for the $\sqrt{s}_{max}$ = 7 TeV ensemble. 
Fit results with initial values
corresponding to the final values
given in Table \ref{tab:1}: Direct Fit and Variants 1, 2 and 3. Units as in Table \ref{tab:1}. }
 \label{tab:2}
 \begin{tabular}{ccccc}
\hline\noalign{\smallskip}
      &  Direct Fit   & V1 & V2  &  V3    \\
\noalign{\smallskip}\hline\noalign{\smallskip}
$a_1$ &  54.9$\pm$5.3 & 57.7$\pm$2.0 & 53.8$\pm$3.0 & 58.4$\pm$2.4  \\
$b_1$ &  0.539$\pm$0.080 & 0.526$\pm$0.010 & 0.452 (fixed) & 0.452 (fixed)  \\
$a_2$ &  33.2$\pm$2.3 & 33.2$\pm$1.7 & 35.77$\pm$0.36 & 35.78$\pm$0.36 \\
$b_2$ &  0.541$\pm$0.016 & 0.541$\pm$0.012 & 0.558 (fixed) & 0.558 (fixed) \\
$\alpha$ &  35.9$\pm$1.9 & 34.90$\pm$0.23 & 32.68$\pm$0.87 & 30.41$\pm$1.00  \\
$\beta$ & 0.290$\pm$0.027 & 0.199$\pm$0.014 & 0.257$\pm$0.017 & 0.093$\pm$0.019  \\
$\gamma$ &  2 (fixed) & 2.104$\pm$0.027 & 2 (fixed) & 2.273$\pm$0.039  \\
$s_h$ &  26$\pm$20 & 11.0$\pm$1.4 & 7.8$\pm$3.5 & 0.77$\pm$0.58 \\\hline
DOF &  157 & 156 & 159 & 158 \\
$\chi^2/$DOF &  0.930 & 0.935 & 0.931 &  0.933 \\
\noalign{\smallskip}\hline
\end{tabular}
\end{table}

\begin{figure}[h]
\centering
\includegraphics[scale=0.7]{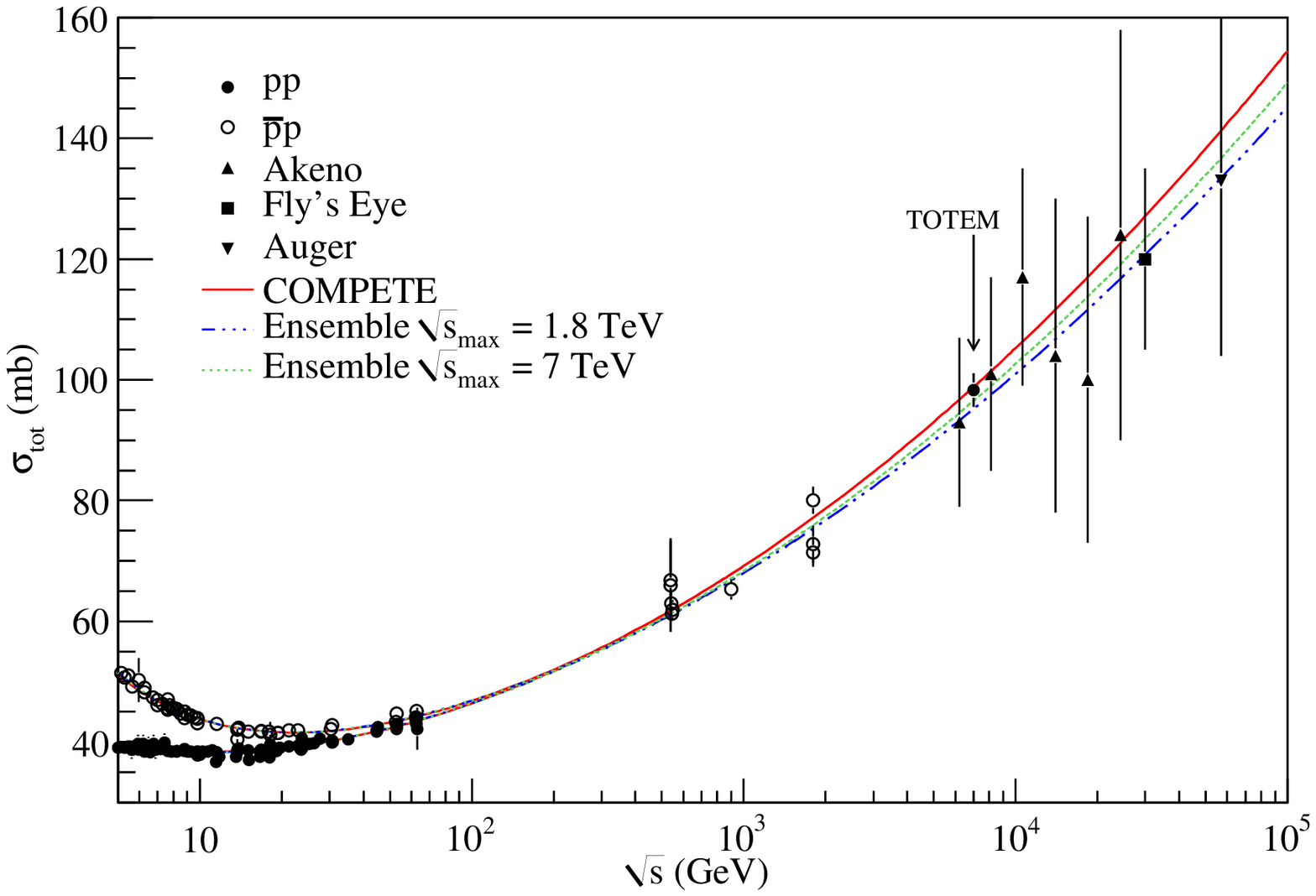} 
\caption{(Color online) Comparison between experimental data
and fit results. The dot-dashed and
dotted lines display the results of the Direct Fit-Method 1 applied
to the $\sqrt{s}_{max}$ = 1.8 TeV and $\sqrt{s}_{max}$ =
7 TeV ensembles, respectively. The solid line is the fit obtained by 
the COMPETE Collaboration. Numerical information displayed in
Table \ref{tab:1} (second and third columns) and Table \ref{tab:2} (second column).}
\label{fig:1}
\end{figure}

\begin{figure}[h]
\centering
\includegraphics[scale=0.7]{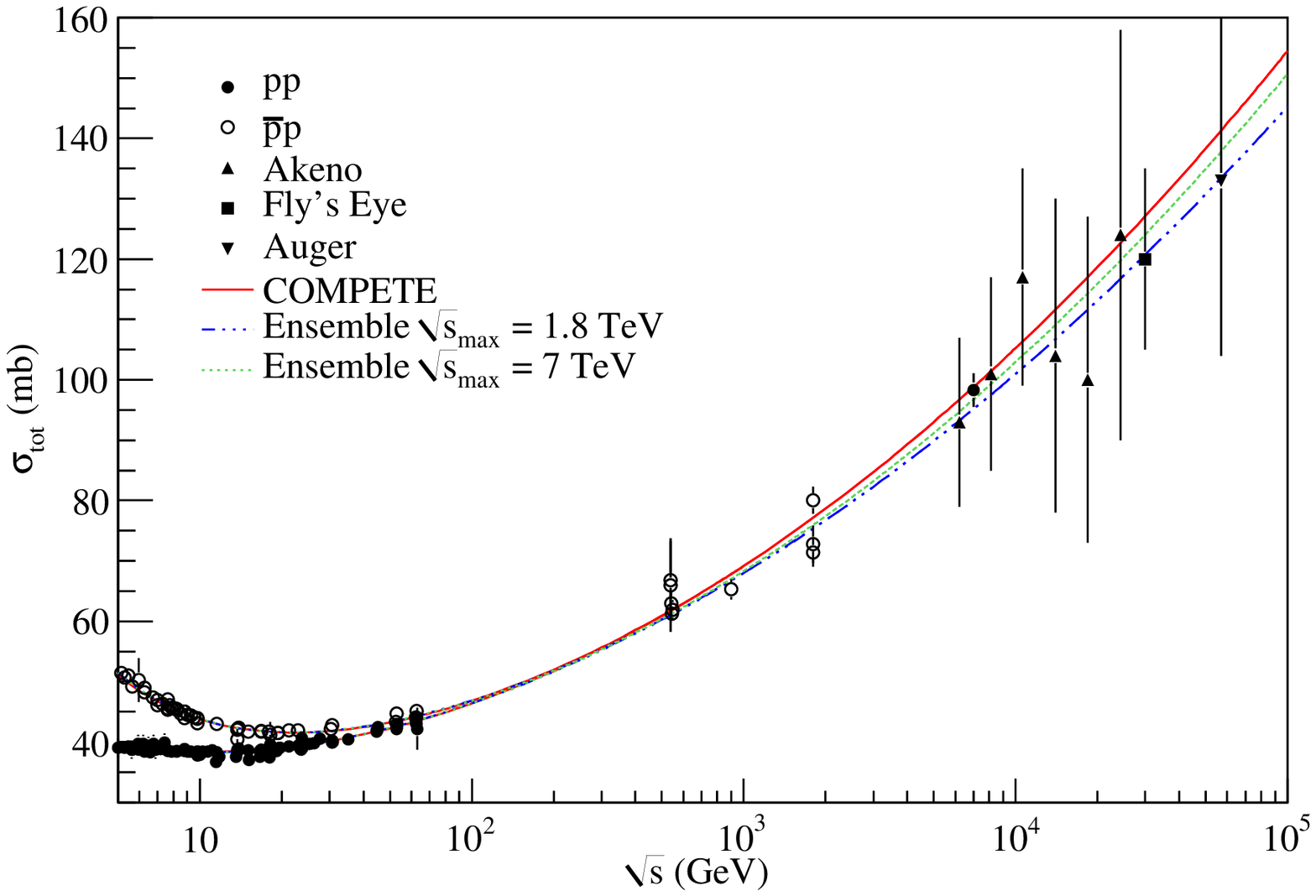}
\caption{(Color online) Analogous to Fig. \ref{fig:1}, the dot-dashed and
  dotted lines computed with Variant 1 of Method 1
  (V1). Numerical information displyed in Tables \ref{tab:1} (fourth column) and \ref{tab:2} 
  (third column).}
\label{fig:2}
\end{figure}

\begin{figure}[h]
\centering
\includegraphics[scale=0.7]{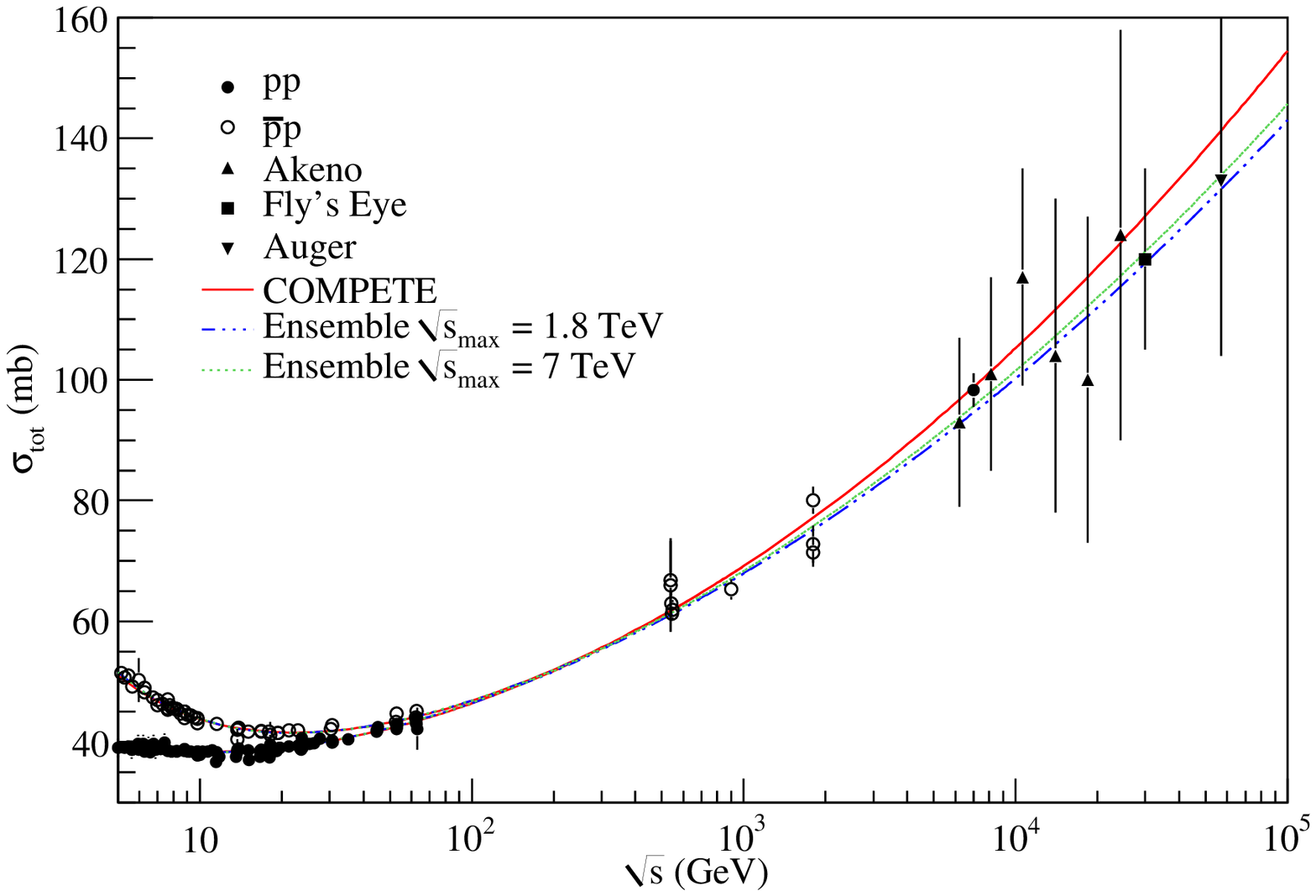}
\caption{(Color online) Analogous to Fig. \ref{fig:1}, the dot-dashed and
  dotted lines computed with Variant 2 of Method 1
  (V2). Numerical information displyed in Tables \ref{tab:1} (fifth column) and \ref{tab:2} 
  (fourth column).}
\label{fig:3}
\end{figure}

\begin{figure}[h]
\centering
\includegraphics[scale=0.7]{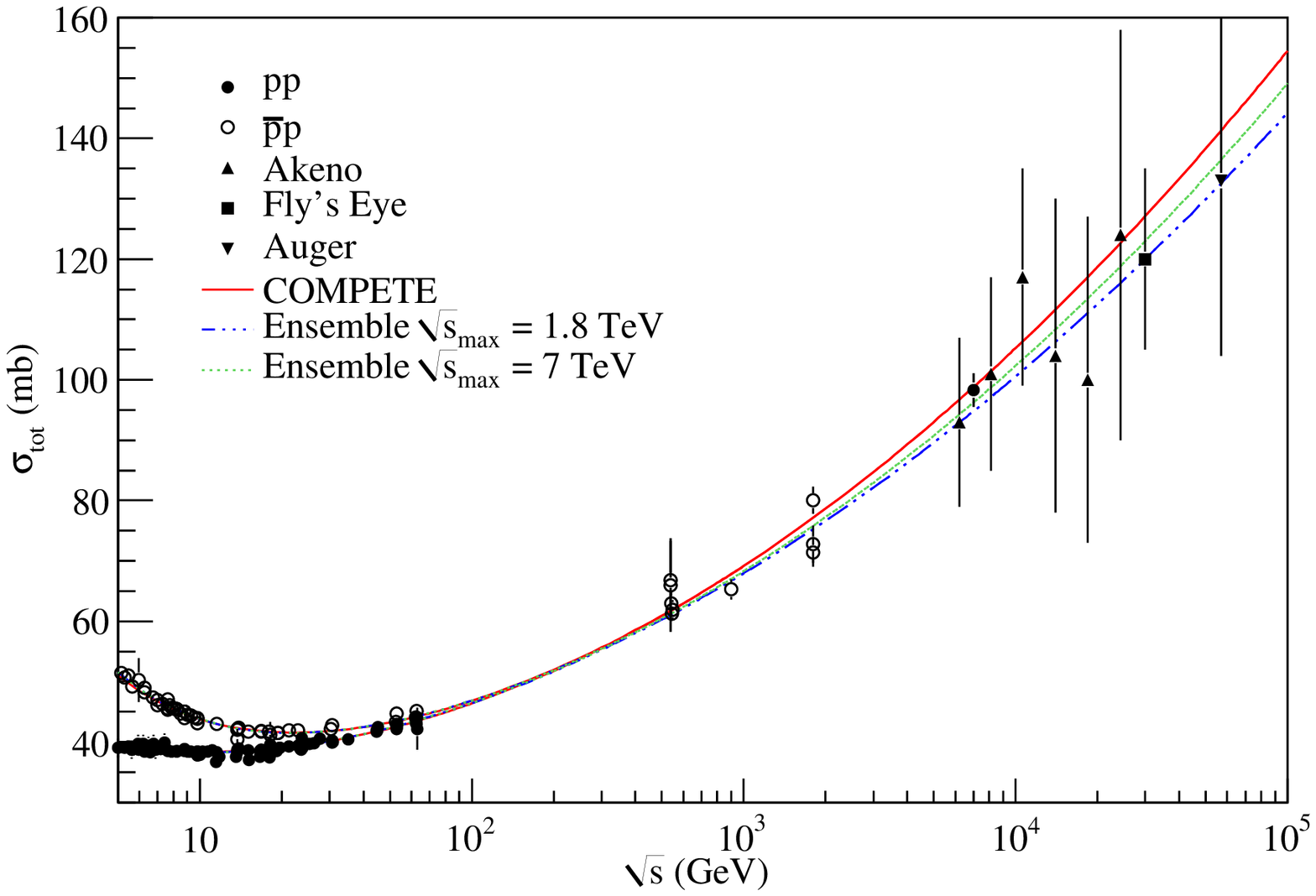}
\caption{(Color online) Analogous to Fig. \ref{fig:1}, the dot-dashed and
  dotted lines computed with Variant 3 of Method 1
  (V3). Numerical information displyed in Tables \ref{tab:1} and \ref{tab:2} (last columns).}
\label{fig:4}
\end{figure}

The corresponding curves obtained with Method 1 for both ensembles and the experimental data are 
shown in Figures \ref{fig:1} to \ref{fig:4}. In each figure the COMPETE result
is displayed as reference (solid lines), together with each result obtained with ensembles 
$\sqrt{s}_{max}$ = 1.8 TeV (dot-dashed lines) and $\sqrt{s}_{max}$ = 7 TeV (dotted lines): Direct Fit and Variants
1, 2 and 3 in Figures \ref{fig:1}, \ref{fig:2}, \ref{fig:3} and \ref{fig:4}, respectively.

In these figures it is also displayed estimates for the total cross
section from cosmic-ray experiments at the highest energies, which were not included in the
data reductions. These estimates,
discussed in \cite{alm} (see also \cite{alm01}), were obtained by 
the Akeno Collaboration \cite{akeno} in the  $\sim$ 6 - 24 TeV region, and by the Fly's Eye Collaboration at 30 TeV \cite{fly}.
Also included is the recent result by the Pierre Auger Collaboration at
57 TeV \cite{auger}.
We shall discuss these and the results that follows in Section \ref{sec:4}.

\subsection{Method 2}
\label{subsec:3.2}

\subsubsection{Results for the $\sqrt{s}_{max}$ = 1.8 TeV Ensemble}
\label{subsubsec:3.2.1}

The initial values, referred to in the beginning of this Section, are displayed in the 
second column of Table \ref{tab:3}. Here,
the following variants have been considered:

\begin{description}

\item[\bf{Variant 4 (V4)}:]

The parameters $b_1$ and $b_2$ have been fixed to the average values expected for
the Reggeon intercepts and $\gamma$ = 2 also fixed (third column in Table \ref{tab:3}).

\item[\bf{Variant 5 (V5)}:]

Same as Variant 4 with $\gamma$ as free parameter
(fourth column in Table \ref{tab:3}).

\item[\bf{Variant 6 (V6)}:]

Initialized with the parameters from Variant 4:  $b_1$, $b_2$ and $\gamma$
free parameters (fifth column in Table \ref{tab:3}).

\end{description}

The following diagram summarizes the fit procedures and initial
values (Table \ref{tab:3}):

\begin{equation}\nonumber
\text{Initial Values} \rightarrow 
\left\{\begin{array}{rl}
\text{Variant\ 4} & \rightarrow \text{Variant 6} \\
\text{Variant\ 5} & \\
\end{array}\right.
\end{equation}

\subsubsection{Results for the $\sqrt{s}_{max}$ = 7 TeV Ensemble}
\label{subsubsec:3.2.2}

Here we followed the same procedure described above, now with the
expanded ensemble. Table \ref{tab:4} displays
the numerical results and statistical
information. Figures \ref{fig:5}, \ref{fig:6} and \ref{fig:7} compare the accelerator data and estimates from cosmic-ray experiments
with the fits obtained with Method 2 and variants 4, 5 and 6, respectively for the
$\sqrt{s}_{max}$ = 1.8 TeV (dot-dashed lines) and $\sqrt{s}_{max}$ = 7 TeV
(dotted lines) ensembles.

\begin{table}[ht]
 \centering
 \caption{Results of Method 2 for the $\sqrt{s}_{max}$ = 1.8 TeV ensemble.
Fit results with Variants 4, 5 and 6. Units as in Table \ref{tab:1}.}
 \label{tab:3}
 \begin{tabular}{ccccc}
\hline\noalign{\smallskip}
 & Initial Values & V4 & V5 & V6 \\
\noalign{\smallskip}\hline\noalign{\smallskip}
$a_1$ & 1 & 52.6$\pm$3.4 & 54.7$\pm$2.6 & 53.8$\pm$1.8 \\
$b_1$ & 0.5 & 0.5 (fixed) & 0.5 (fixed) & 0.4952$\pm$0.0099 \\
$a_2$ & 1 & 27.70$\pm$0.28 & 27.71$\pm$0.28 & 33.1$\pm$1.7 \\
$b_2$ & 0.5 & 0.5 (fixed) & 0.5 (fixed) & 0.540$\pm$0.012 \\
$\alpha$ & 1 & 34.86$\pm$0.69 & 34.32$\pm$0.77 & 34.60$\pm$0.24 \\
$\beta$ & 50 & 0.270$\pm$0.019 & 0.21$\pm$0.11 & 0.290$\pm$0.021 \\
$\gamma$ & 2 & 2 (fixed) & 2.06$\pm$0.16 & 1.975$\pm$0.028 \\
$s_h$ & 1 & 15.6$\pm$6.3 & 9.3$\pm$8.4 & 16.0$\pm$2.0 \\
\noalign{\smallskip}\hline
DOF & - & 158 & 157 & 155 \\
$\chi^2/$DOF & - & 0.963 & 0.969 & 0.937 \\
 \hline
\end{tabular}
\end{table}

\begin{table}[ht]
 \centering
 \caption{Results of Method 2 for the $\sqrt{s}_{max}$ = 7 TeV ensemble. 
Fit results with Variants 4, 5 and 6. Same units as in Table \ref{tab:1}.}
 \label{tab:4}
  \begin{tabular}{cccc}
\hline\noalign{\smallskip}
 & V4 & V5 & V6 \\
\noalign{\smallskip}\hline\noalign{\smallskip}
$a_1$ & 51.34$\pm$3.2 & 56.5$\pm$1.1 & 57.6$\pm$7.5 \\
$b_1$ & 0.5 (fixed) & 0.5 (fixed) & 0.525$\pm$0.077 \\
$a_2$ & 27.70$\pm$0.28 & 27.70$\pm$0.28 & 33.2$\pm$2.3 \\
$b_2$ & 0.5 (fixed) & 0.5 (fixed) & 0.541$\pm$0.016 \\
$\alpha$ & 35.13$\pm$0.60 & 33.65$\pm$0.22 & 34.9$\pm$1.7 \\
$\beta$ & 0.279$\pm$0.016 & 0.1301$\pm$0.0086 & 0.199$\pm$0.064 \\
$\gamma$ & 2 (fixed) & 2.213$\pm$0.024 & 2.10$\pm$0.11 \\
$s_h$ & 18.5$\pm$6.3 & 3.90$\pm$0.52 & 10.8$\pm$5.6 \\\hline
DOF & 159 & 158 & 156 \\
$\chi^2/$DOF & 0.962 & 0.967 & 0.935 \\
\noalign{\smallskip}\hline
\end{tabular}
\end{table}

We note a difference in the diagrams related to initial values and variants 
(two schemes displayed in this Section). This effect is
consequence of fit procedures since in certain cases the data reductions led back 
to the initial values
or to solutions with errors that exceded the central values of the parameters.
All these cases have been excluded.

\section{Conclusions and Final Remarks}
\label{sec:4}

We have employed two  methods to initialize the fitting 
parameters, six variants and two data ensembles. 
The results for the $\sqrt{s}_{max}$ = 1.8 TeV ensemble,
listed in Tables \ref{tab:1} and \ref{tab:3}, are statistically consistent
with $\gamma$ = 2, which points to a saturation
of the squared logarithm dependence for the total cross section at high energies.
More specifically, the results can be summarized as follows:

\begin{description}

\item[] Method 1 (Table \ref{tab:1}): $\gamma \approx$ 2.00 $\pm$ 0.03 (V1) and $\gamma \approx$ 2.08 $\pm$ 0.10 (V3).

\item[] Method 2 (Table \ref{tab:3}): $\gamma \approx$ 2.06 $\pm$ 0.16 (V5) and $\gamma \approx$ 1.98 $\pm$ 0.03 (V6).

\end{description}

On the other hand, when the more recent 7 TeV TOTEM data is included,
all fits with $\gamma$ as a free parameter, listed in Tables \ref{tab:2} and \ref{tab:4},
are statistically consistent with exponents above 2, as indicated in the following summary:

\begin{description}

\item[] Method 1 (Table \ref{tab:2}): $\gamma \approx$ 2.10 $\pm$ 0.03 (V1) and $\gamma \approx$ 2.27 $\pm$ 0.04 (V3).

\item[] Method 2 (Table \ref{tab:4}): $\gamma \approx$ 2.21 $\pm$ 0.02 (V5) and $\gamma \approx$ 2.10 $\pm$ 0.11 (V6).

\end{description}

All these results for $\gamma$, together with those obtained by Amaldi \textit{et al}. and the
UA/4 Collaboration, are schematically displayed in Figure \ref{fig:8}.
From our data reductions through parametrization (\ref{eq:3}-\ref{eq:5}) to $pp$ and $\bar{p}p$ scattering above
5 GeV, including the 7 TeV TOTEM result, we conclude that the total hadronic cross 
section may rise faster than $\ln^2{s}$ at high energies.

From Figure \ref{fig:8}, we see that
the central values of the $\gamma$ parameter from the fits with 
$\sqrt{s}_{max}$ = 1.8 TeV (Methods 1 and 2) are below the corresponding
values obtained by Amaldi \textit{et al}. and by the UA4/2 Collaboration. Since these authors
have included the $\rho$ information in their analysis, it seems not clear that
simultaneous fits to  $\rho$ and $\sigma_{tot}$ data may, in that case, anchor the
rise of the total cross section. In other words, we understand that the
results we have obtained with $\sqrt{s}_{max}$ = 1.8 TeV and 
$\sqrt{s}_{max}$ = 7 TeV ensembles are not connected with the inclusion or not
of the $\rho$ information, but are direct consequences of the total cross section data
analyzed and the possibility to treat $\gamma$ as a free fit parameter (not fixed to 2).
In fact, our \textit{predictions} for $\rho(s)$ in the case of $\gamma >\ $2 are presented in
Appendix \ref{app:A} and show good agreement with experimental information available above 5 GeV.

Concerning the rise of the total cross section faster than the squared logarithm 
dependence, we have the following final comments:

1. In the theoretical context, different assumptions on the nonphysical
amplitudes in the asymptotic region can explain this behavior, which
does not mean violation of unitarity, as recently discussed 
by Azimov \cite{azimov}.

\begin{figure}[h]
\centering
\includegraphics[scale=0.7]{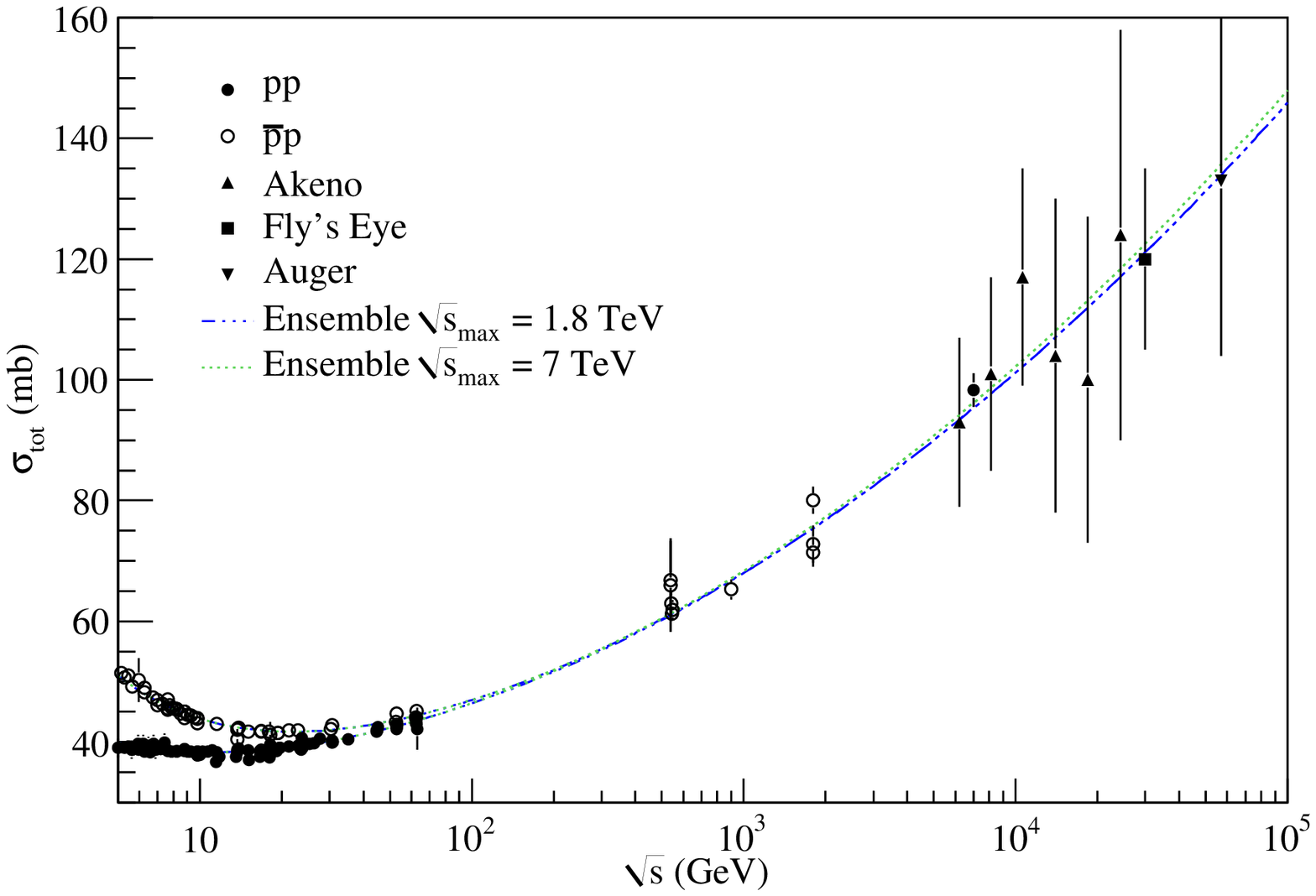}
\caption{(Color online) Analogous to Fig. \ref{fig:1}, the dot-dashed and
  dotted lines computed with Variant 4 of Method 2
  (V4). Numerical information displayed in Tables \ref{tab:3} (third
  column) and \ref{tab:4} (second column).}
\label{fig:5}
\end{figure}

\begin{figure}[h]
\centering
\includegraphics[scale=0.7]{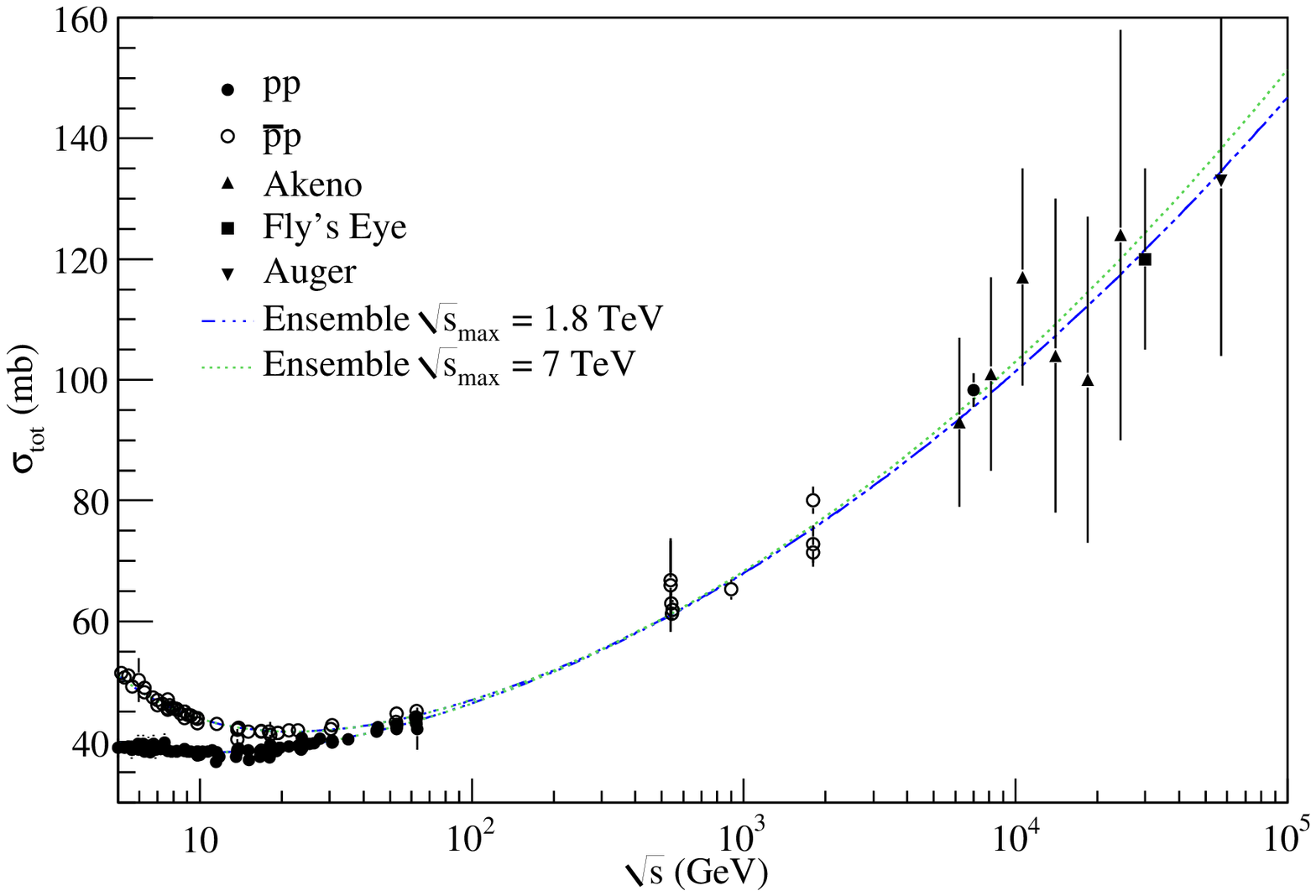}
\caption{(Color online) Analogous to Fig. \ref{fig:1}, the dot-dashed and
  dotted lines computed with Variant 5 of Method 2
  (V5). Numerical information displyed in Tables \ref{tab:3} (fourth
  column) and \ref{tab:4} (third column).}
\label{fig:6}
\end{figure}

\begin{figure}[h]
\centering
\includegraphics[scale=0.7]{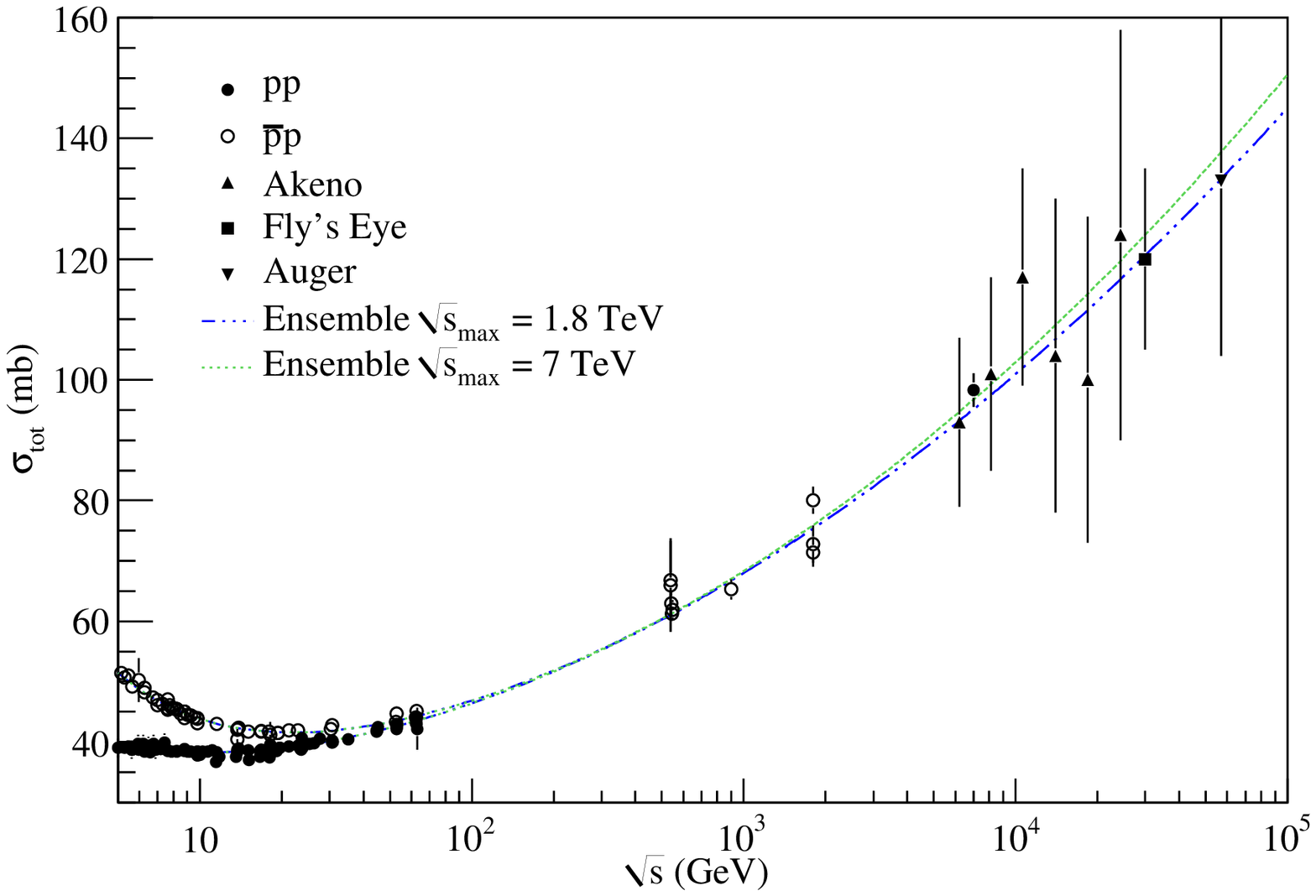}
\caption{(Color online) Analogous to Fig. \ref{fig:1}, the dot-dashed and
  dotted lines computed with Variant 6 of Method 2
  (V6). Numerical information displayed in Tables \ref{tab:3} (fifth
  column) and \ref{tab:4} (fourth column).}
\label{fig:7}
\end{figure}

\begin{figure}[h]
\centering
\includegraphics[scale=0.7]{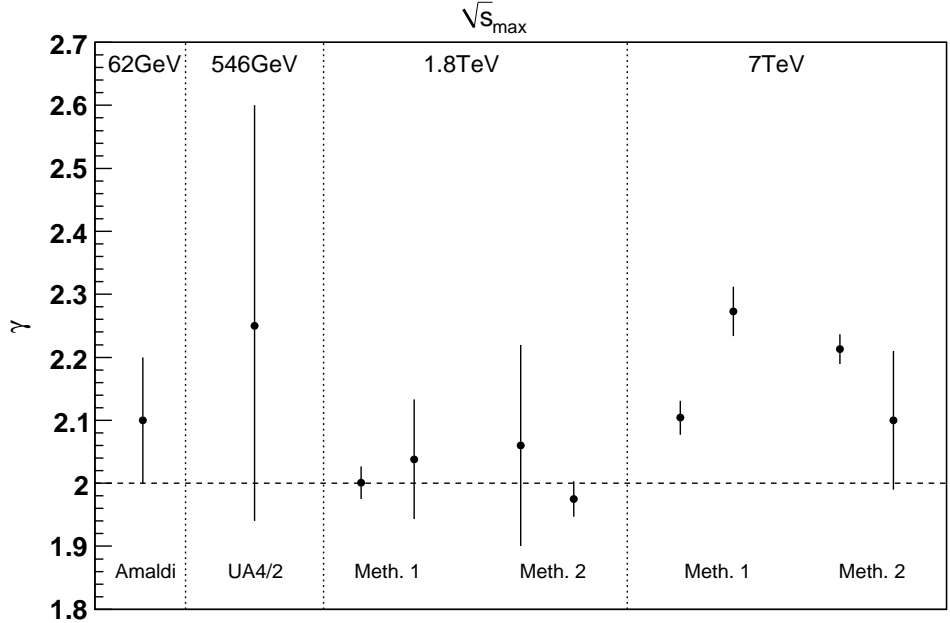}
\caption{Results for the exponent $\gamma$ as a free parameter in different
data reductions through parametrization (\ref{eq:3}-\ref{eq:5}). Shown are the
  results by Amaldi \emph{et al.} \cite{amaldi} ($\sqrt{s}_{max}$ =
  62 GeV), the UA4/2 Collaboration \cite{ua42} ($\sqrt{s}_{max}$ =
  546 GeV), and from our analyses for the
  $\sqrt{s}_{max}$ = 1.8 TeV and $\sqrt{s}_{max}$ =
  7 TeV ensembles with Methods 1 and 2.}
\label{fig:8}
\end{figure}

2. In the experimental context, beyond the LHC energy region, 
extremely large uncertainties are associated with estimations of the $pp$
total cross section from cosmic-ray experiments. Such uncertainties are unavoidable
since the flux decreases as the energy grows and because the extraction
of $\sigma_{pp}$ from $\sigma_{p-air}$ is model dependent. These estimations have been included
in our figures for qualitative comparison only. It may be useful to recall that the 
the results of the Akeno Collaboration \cite{akeno} 
have been criticized in several works, as discussed for example in Refs. \cite{compete2} and
\cite{alm}, and references therein. 
Had we ignored this information, the Fly's Eye and Auger
results, i.e., the highest-energy points in our figures, might suggest 
a different scenario for the
rise of the total cross section. 
In fact, from Figures \ref{fig:1} to \ref{fig:7}, all the curves in consistency with the TOTEM
data lie above the central values of these cosmic-ray estimations and the same is true in the
inverse sense. That is, \textit{there is no agreement among the TOTEM result and the
Fly's Eye and Auger estimations, at least in what concerns parametrization} (\ref{eq:3}-\ref{eq:5}). In this sense,
the expected measurement of $\sigma_{tot}$ by the TOTEM Collaboration at 7 and 8 TeV through the luminosity-independent \cite{pred,matthiae}
method may shed light on the subject. Moreover,
new experimental results on elastic and diffractive scattering at 8 TeV (and, subsequently, at 14 TeV),
will provide novel phenomenological insights and reduce
the uncertainties from model extrapolations
necessary to obtain  $\sigma_{tot}$ from cosmic-ray experiments (see \cite{secb}
and references therein).

At last we stress that our approach and strategies do not follow the usual or standard
lines of the amplitude analyses on the energy dependence of the total cross section. 
As commented, the latter include
$\rho$ data, associate the parametrization with data at low energies and constrain
the fits to describe other reactions, for which low-energy data are available.
These are aspects that we expect to consider in a future work, since 
they may provide information that is complementary to the results we have presented.
As commented along the manuscript,
our goal has been to explore the possibility of investigating
only the reactions with highest energy interval available, concentrating
on the total cross section data. We have tried to identify possible high-energy effects
that may be unrelated to the trends of
the lower-energy data (including those from other reactions) or hidden in fits
including the $\rho$ information. 
Our intention is not to compete with other authors or analysis,
but to call the attention to the possibility that
the rise of the total hadronic cross section may still constitute
an open problem, an assertion that contrasts with the view advocated by some authors \cite{bh3}.

\vspace{0.3cm}

\textit{Note added during revision}

After this paper was submitted to publication, new fits 
of forward quantities were obtained by the Particle Data Group \cite{pdg2012}.
The updated data set includes the TOTEM result at 7 TeV and the new fit with
the highest rank COMPETE parametrization ($\gamma  = 2$) shows that
the point at 7 TeV is not described: the curve lies below the data (Figure 46.10 in \cite{pdg2012}). This result corroborates those already shown with the
$\sqrt{s}_{max}$ = 7 TeV ensemble in our Figures
\ref{fig:1}, \ref{fig:3} and \ref{fig:5}.

\begin{acknowledgments}
We are thankful to an anonymous referee for valuable comments.
This research was supported by the FAPESP 
under contracts 11/15016-4, 11/00505-0 and 09/50180-0. 
\end{acknowledgments}

\appendix
\section{Predictions for the ratio between the real and imaginary parts of the forward amplitude}
\label{app:A}

In this appendix we present  predictions for $\rho(s)$ 
from fits to $pp$ and $\bar{p}p$ total cross section data through
parametrization (\ref{eq:3}-\ref{eq:5}) with $\gamma > 2$. We also discuss the effect and role of the subtraction
constant, embodied in the associated dispersion relation. In search of extreme
effects we shall consider the two 
results obtained with the $\sqrt{s}_{max}$ = 7 TeV ensemble,
which correspond to
the highest exponents $\gamma$,
namely

\begin{description}

\item[] Method 1 - Variant 3: $\gamma \approx$ 2.27 $\pm$ 0.04 (Table \ref{tab:2})

\item[] Method 2 - Variant 5: $\gamma \approx$ 2.21 $\pm$ 0.02 (Table \ref{tab:4})

\end{description}

In terms of the \textit{forward} scattering amplitude $F(s)$,
the total cross section (optical
theorem) and the ratio between the real and imaginary parts of the amplitude,
at high energies, can be expressed by

\begin{eqnarray} 
\sigma_{tot}(s) = \frac{\textrm{Im}\,F(s)}{s}, 
\qquad
\rho(s) = \frac{\textrm{Re}\,F(s)}{\textrm{Im}\,F(s)}. 
\label{eq:6}
\end{eqnarray} 
For crossing even ($+$) and odd ($-$) amplitudes, $pp$ and $\bar{p}p$ scattering
are related by 

\begin{eqnarray}  
F_{\pm}(s) = \frac{F_{pp} \pm 
F_{\bar{p}p}}{2}
\label{eq:7}
\end{eqnarray} 
and it follows from Eqs. (\ref{eq:6}) and  (\ref{eq:7}) that 

\begin{eqnarray}
\rho ^{pp} = \frac{1}{\sigma_{tot}^{pp} (s)} 
\left\{ \frac{\textrm{Re}\,F_{+}}{s}  + \frac{\textrm{Re}\,F_{-}}{s} \right\}
\label{eq:8}
\end{eqnarray}
and

\begin{eqnarray}
\rho ^{\bar{p}p} = \frac{1}{\sigma_{tot}^{\bar{p}p}} 
\left\{ \frac{\textrm{Re}\,F_{+}}{s}  - \frac{\textrm{Re}\,F_{-}}{s} \right\}.
\label{eq:9}
\end{eqnarray}

The real and imaginary parts of the even and odd amplitudes are connected by
dispersion relations and the high-energy domain demands one subtraction \cite{disprel1,disprel2}.
Since we are looking for analytical results, we shall work here with the derivative dispersion relations
in the standard form deduced by Bronzan, Kane and Sukhatme \cite{bks}.
They are obtained from the integral
dispersion relations in the high-energy limit: 

\begin{eqnarray} 
\frac{\textrm{Re}\ F_{+}(s)}{s} = 
\frac{K}{s} + 
\tan\left[\frac{\pi}{2}\frac{d}{d\ln s} \right] 
\frac{\textrm{Im}\ F_{+}(s)}{s}, \nonumber
\end{eqnarray}

\begin{eqnarray} 
\frac{\textrm{Re}\ F_{-}(s)}{s} =
\tan\left[\frac{\pi}{2}\left( 1 + \frac{d}{d\ln s}\right) \right]
\frac{\textrm{Im}\ F_{-}(s)}{s}. \nonumber
\end{eqnarray} 
where $K$ is the \textit{subtraction constant}.
Operationally these relations can be evaluated through the expansions \cite{kn,am}

\begin{eqnarray} 
\frac{\textrm{Re}\ F_{+}(s)}{s} - \frac{K}{s} & = &
\left[ \frac{\pi}{2} \frac{d}{d\ln s} + 
\frac{1}{3} \left(\frac{\pi}{2}\frac{d}{d \ln s}\right)^3 + \frac{2}{15} \left(\frac{\pi}{2}\frac{d}{d \ln s}\right)^5 + \dots \right] \frac{\textrm{Im}\ F_{+}(s)}{s},
\label{eq:10}
\end{eqnarray}

\begin{eqnarray} 
\frac{\textrm{Re}\ F_{-}(s)}{s}  &=& 
- \int \left\{ \frac{d}{d\ln s} \left[\cot \left( \frac{\pi}{2} 
\frac{d}{d\ln s} \right)\right]\frac{\textrm{Im}\ F_{-}(s)}{s} \right\} d\ln s \nonumber \\
&=&
- \frac{2}{\pi}\int \left\{ \left[ 1 - \frac{1}{3} \left(\frac{\pi}{2}\frac{d}{d \ln
s}\right)^2 - \frac{1}{45} \left(\frac{\pi}{2}\frac{d}{d \ln s}\right)^4
 - \dots \right] \frac{\textrm{Im}\ F_{-}(s)}{s} \right\} d \ln s.
\label{eq:11}
\end{eqnarray} 

With parametrization (\ref{eq:3}-\ref{eq:5}) taken as input,
a closed form results from the sum of the contributions
associated with the power-law term $\sigma_{LE}$.
The sum of the contributions from the
logarithm term $\sigma_{HE}$ converges fast for $\gamma$
in the range (2.2 - 2.3) and a third-order approximation
is therefore sufficient. We obtain for the even part

\begin{eqnarray} 
\frac{\textrm{Re}\, F_{+}(s)}{s} & = & \frac{K}{s} 
- a_1 \tan \left(\frac{\pi\, b_1}{2}\right) \left[\frac{s}{s_l}\right]^{-b_1} +
A\ln^{\gamma - 1}\left(\frac{s}{s_h}\right) + \nonumber \\
& & B\ln^{\gamma - 3}\left(\frac{s}{s_h}\right) +
C\ln^{\gamma - 5}\left(\frac{s}{s_h}\right),
\label{eq:12}
\end{eqnarray} 
where
\begin{eqnarray} 
A &=& \frac{\pi}{2} \, \beta\, \gamma,  \qquad \qquad
B = \frac{1}{3} \left[\frac{\pi}{2}\right]^3 \, \beta\, \gamma\, [\gamma - 1][ \gamma - 2], \nonumber \\ 
                                                                 \nonumber \\
C &=& \frac{2}{15} \left[\frac{\pi}{2}\right]^5 \, \beta\, \gamma\, [\gamma - 1][ \gamma - 2]
[\gamma - 3][ \gamma - 4]
\label{eq:13}
\end{eqnarray} 
and for the odd part

\begin{eqnarray} 
\frac{\textrm{Re}\, F_{-}(s)}{s} =
-\, a_2\, \cot \left(\frac{\pi\, b_2}{2}\right) \left[\frac{s}{s_l}\right]^{-b_2}.
\label{eq:14}
\end{eqnarray} 

With the above results, Eqs. (\ref{eq:8}) and (\ref{eq:9}) yield analytical expressions for $\rho ^{pp}(s)$ and
$\rho ^{\bar{p}p}(s)$. Note that the analytical results imply that as $s \rightarrow \infty$: 
\begin{eqnarray} 
\rho \ \propto \ 1/\ln s \ \rightarrow \ 0.
\nonumber
\end{eqnarray}

\begin{figure}[h]
\includegraphics[scale=0.7]{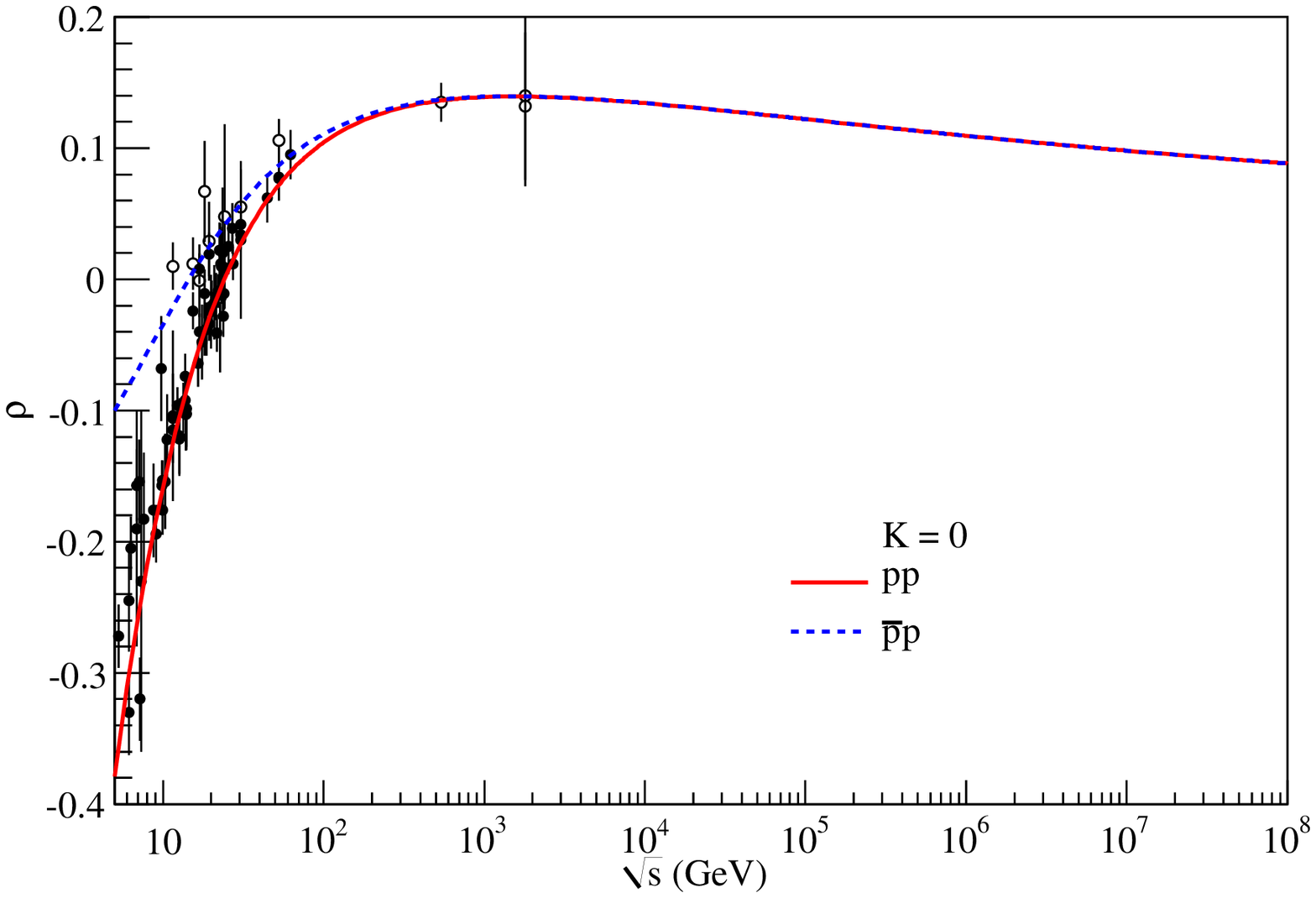}\\
\includegraphics[scale=0.7]{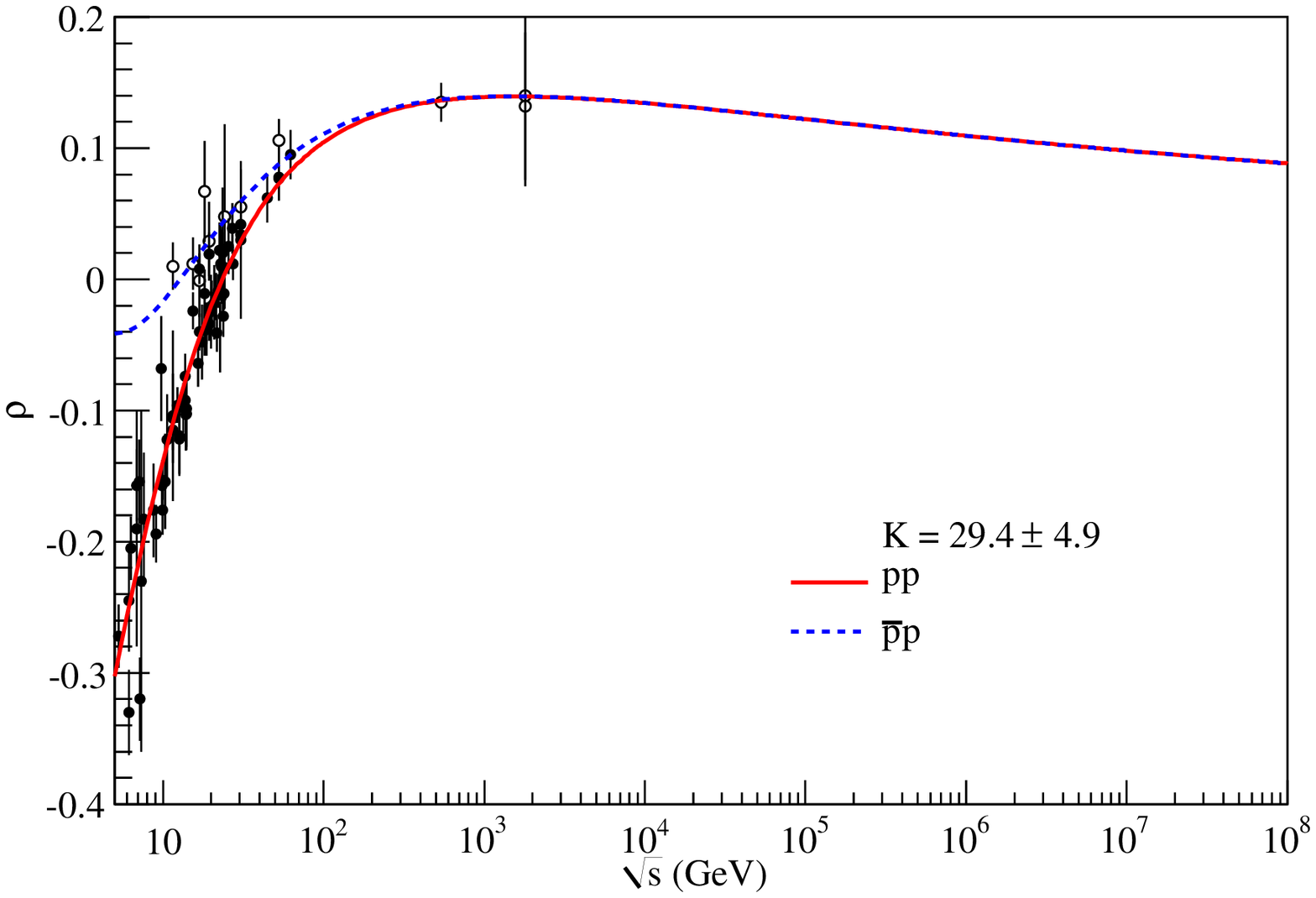}
\caption{(Color online) Comparison between experimental results
  for $\rho$ and the analytical expressions derived in Appendix \ref{app:A},
  with the parameters obtained by Variant 3 of Method 1, listed in
  Table \ref{tab:2}, which yield $\gamma\approx 2.27\pm0.04$. The filled and
  the open circles represent the experimental data for $pp$ and $\bar
  p p$ scattering respectively. The top panel shows our predictions
  with $K=0$, which correspond to $\chi^{2}/DOF=1.93$. The bottom
  panel, is the best fit to the $\rho$ data with $K$ a free parameter
  ($\chi^{2}/DOF= 1.45$).}
\label{fig:9}
\end{figure}

\begin{figure}[h]
\includegraphics[scale=0.7]{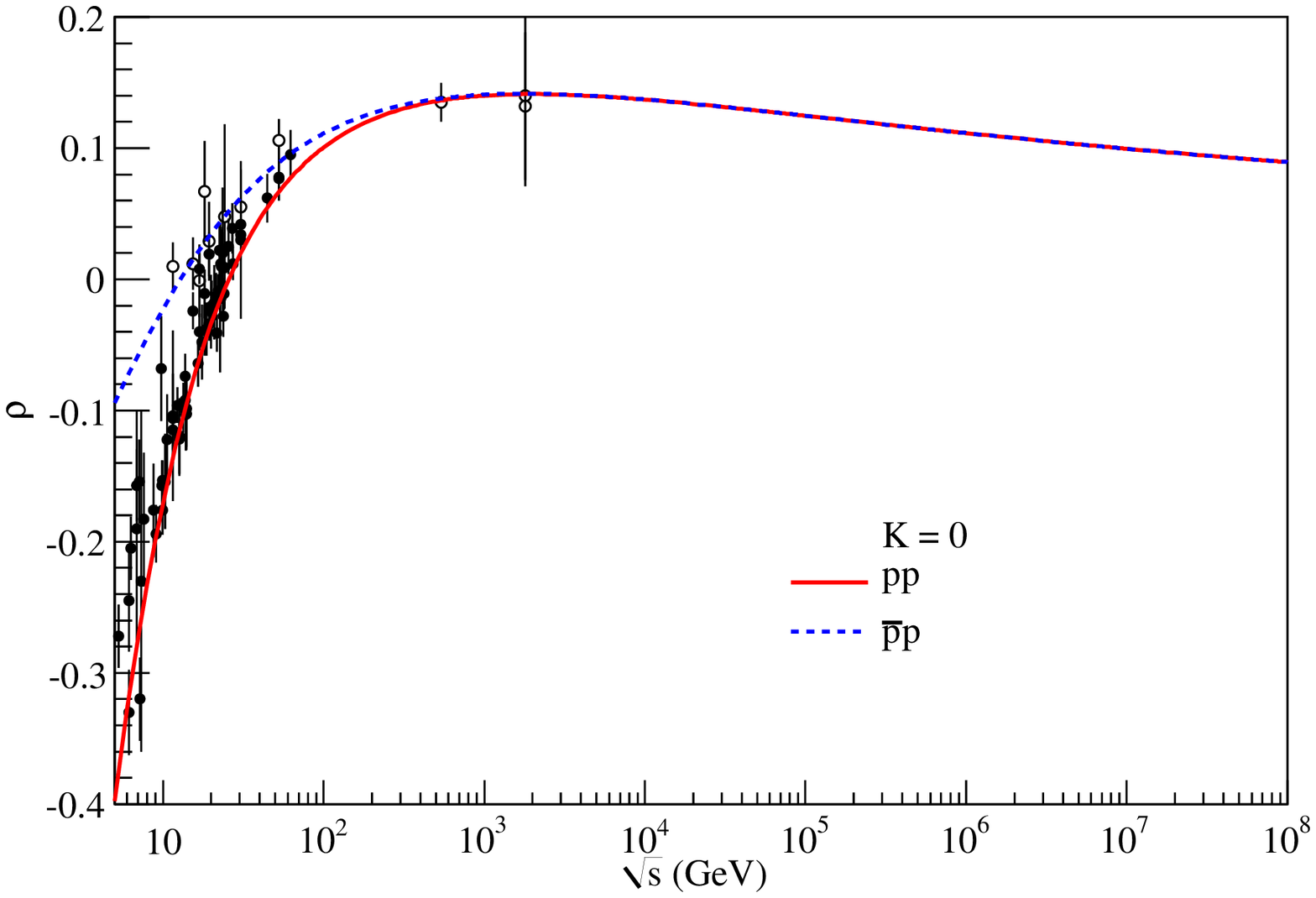}\\
\includegraphics[scale=0.7]{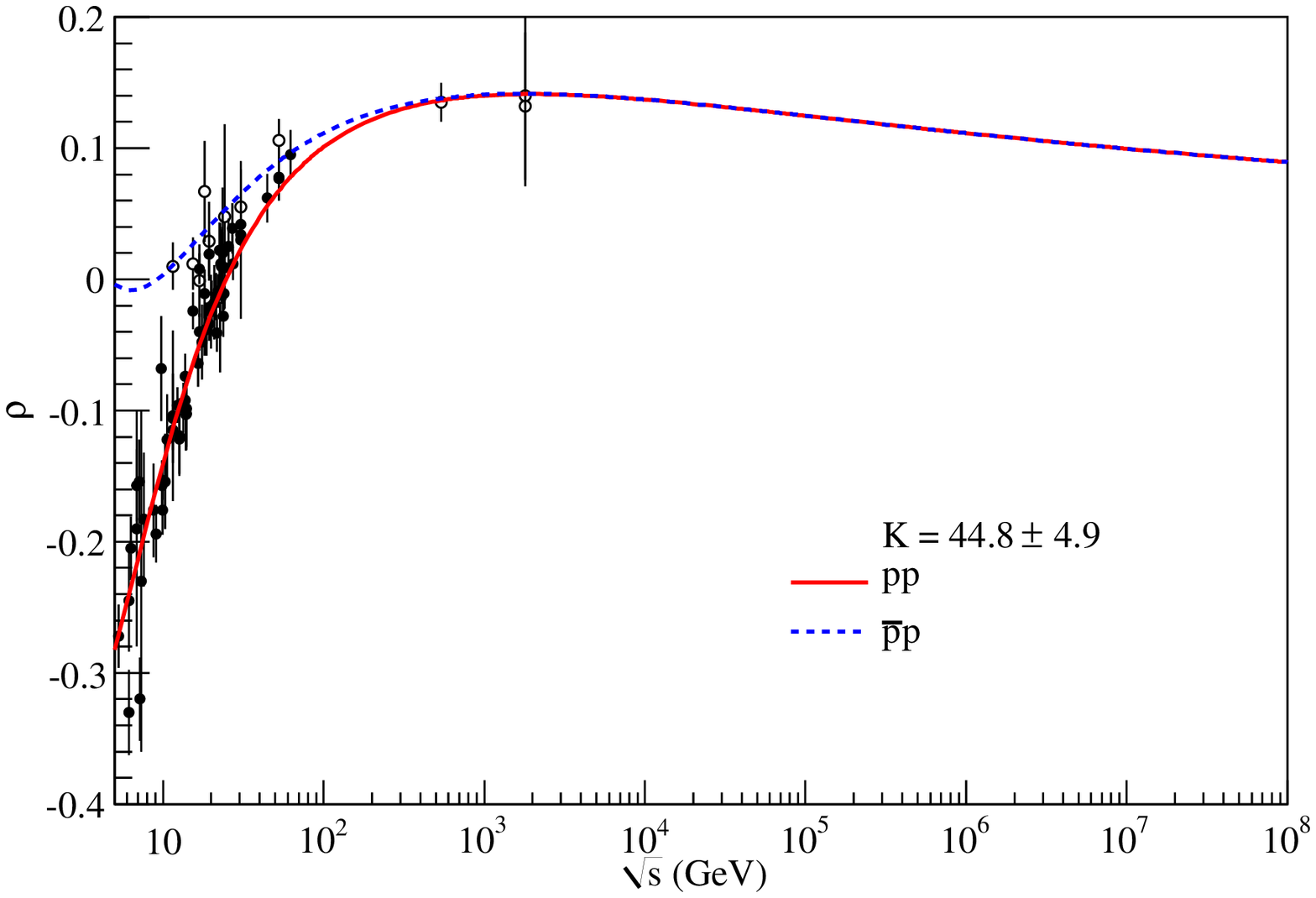}
\caption{(Color online) Analogous to Fig. \ref{fig:9}, the analytical
  results plotted for the parameters obtained with Variant 5 of
  Method 2 (Table \ref{tab:4}), which yield $\gamma\approx 2.21\pm 0.02$. The
  ratios $\chi^{2}/DOF$ in the top and bottom panels are 2.70 and
  1.58, respectively.}
\label{fig:10}
 \end{figure}

In what follows we present the predictions for $\rho(s)$
and compare the results with the experimental information available \cite{pdg2010}.
To study the effect of the subtraction constant, we follow two alternatives:
in the first we fix it at $K = 0$; in the second we let $K$ be a fit parameter to the $\rho$ data only.
More specifically, we take as input the parameters in 
Tables \ref{tab:2} and  \ref{tab:4}, with $s_l$ = 1 GeV$^2$. In the first alternative, 
$K=0$,  we obtain the direct prediction for $\rho(s)$. In the second alternative
we fix all the other parameters and adjust $K$ to fit
the $\rho$ data.
In the latter case, the first run of the MINUIT Code yields the
$\chi^2$/DOF for $K=0$.
The predictions with the results by Method 1 - Variant 3 (Table \ref{tab:2}) are displayed in Figure
\ref{fig:9} and those with the results by  Method 2 - Variant 5 (Table \ref{tab:4}) in Figure \ref{fig:10}.
We are led to the following conclusions:

\begin{description}

\item[1.]
Even in these extrema cases, with $\gamma: 2.2 - 2.3$, all the experimental information on $\rho ^{pp}(s)$ and 
$\rho ^{\bar{p}p}(s)$ above 5 GeV are quite well described;

\item[2.] The subtraction constant affects the low-energy results,
below $\sqrt{s} \sim$ 20 GeV;

\item[3.] The best $\chi^2$/DOF results (closest to 1) are obtained with $K$ as a free fit parameter.
Obviously, that is a consequence of adding one more free parameter, however without
the physical interpretation associated
with  those present in the parametrization of the total cross section. 

\end{description}

Finally, we recall that in simultaneous fit
to $\sigma_{tot}$ and $\rho$  the subtraction constant affects both the low- and 
high-energy regions
\cite{alm,alm01}.
That is a consequence of the strong correlation among the subtraction constant and all
the other physical free fit parameters. We plan to discuss this consequence
and other aspects of the fit procedures in a
forthcoming paper.


\end{document}